\documentclass[preprint,11pt]{aastex}
\slugcomment{{\it to appear in The Astrophysical Journal}}

\newcommand{\lsun}{\hbox{L$_\odot$}}
\newcommand{\msun}{\hbox{M$_\odot$}}
\newcommand{\kms}{\hbox{km s$^{-1}$}}

\newcommand{\ha}{\hbox{H$\alpha$}}
\newcommand{\hb}{\hbox{H$\beta$}}

\newcommand{\hii}{\ion{H}{2}}
\newcommand{\cii}{\ion{C}{2}]}
\newcommand{\nii}{[\ion{N}{2}]}
\newcommand{\sii}{[\ion{S}{2}]}
\newcommand{\oi}{[\ion{O}{1}]}
\newcommand{\oii}{[\ion{O}{2}]}
\newcommand{\oiii}{[\ion{O}{3}]}

\begin{document}

\title{THE CUSPY LINER NUCLEUS OF THE S0/a GALAXY NGC~2681\altaffilmark{1}}

\altaffiltext{1}{Based on observations with the NASA/ESA Hubble Space
Telescope, obtained at the Space Telescope Science Institute, which is
operated by AURA, Inc., under NASA Contract NAS 5-26555.}

\author{Michele~Cappellari\altaffilmark{2,3}, 
Francesco~Bertola\altaffilmark{2}, 
David~Burstein\altaffilmark{4}, 
Lucio~M.~Buson\altaffilmark{5}, 
Laura~Greggio\altaffilmark{6,7}, 
Alvio~Renzini\altaffilmark{3}}

\altaffiltext{2}{Dipartimento di Astronomia,
    Universit\`a di Padova, Padova, Italy}
\altaffiltext{3}{European Southern Observatory,
    Garching bei M\"unchen, Germany}
\altaffiltext{4}{Department of Physics \& Astronomy,
    Arizona State University, Tempe, AZ, USA}
\altaffiltext{5}{Osservatorio di Capodimonte, Napoli, Italy}
\altaffiltext{6}{Dipartimento di Astronomia,
    Universit\`a di Bologna, Bologna, Italy}
\altaffiltext{7}{Sternwarte der Universit\"at M\"unchen,
    M\"unchen, Germany}

\begin{abstract}

The nucleus of the bulge-dominated, multiply-barred S0/a galaxy NGC~2681 is studied in detail, using high resolution Hubble Space Telescope FOC and NICMOS imaging and FOS spectroscopy. The ionised gas central velocity dispersion is found to increase by a factor $\approx2$ when narrowing the aperture from $R\approx1\farcs5$ (ground) to $R\approx0\farcs1$ (FOS). Dynamical modeling of these velocity dispersions suggests that NGC~2681 does host a supermassive black hole (BH) for which one can estimate a firm mass upper limit $M_\bullet\lesssim6\times 10^7$\msun. This upper limit is consistent with the relation between the central BH mass and velocity dispersion $M_\bullet-\sigma$ known for other galaxies. The emission line ratios place the nucleus of NGC~2681 among LINERs. It is likely that the emission line region comes from a rather mild, but steady, feeding of gas to the central BH in this galaxy. The inner stellar population lacks any measurable color gradient (to a radius of 0.6 kpc) from the infrared to the ultraviolet, consistently with FOC, FOS and IUE data, all indicating that this system underwent a starburst $\approx1$ Gyr ago that encompassed its whole interior, down to its very center.  The most likely source of such a widely-distributed starburst is the dumping of tidally-extruded gas from a galaxy neighbor.  If so, then NGC~2681 can be considered as the older brother of M82, seen face-on as opposed to the edge-on view we have for M82.

\end{abstract}


\section{INTRODUCTION\label{sec:intro}}

Hubble Space Telescope (HST), Faint Object Camera (FOC) images for the central regions of the S0/a galaxy NGC~2681 were obtained in four ultraviolet bands in 1993, in the context of a project aimed at understanding the origin of UV emission from bulge-dominated galaxies. A point-like source---which we term {\it spike} for short---was evident at its center with a photometric profile being indistinguishable from the PSF of the aberrated, pre-COSTAR HST images \citep{ber95}.

Motivated by the related discovery of the variable nature of the UV-bright spike in NGC~4552, whose flare has been interpreted as an accretion event onto a central supermassive black hole (BH) \citep{ren95,cap99}, we obtained additional FOC imaging and Faint Object Spectrograph (FOS) spectra of the NGC~2681 spike in 1997, this time with the COSTAR-corrected optics of HST. Unlike the case of NGC~4552, this latter set of HST observations did not reveal any variability of the central spike in NGC~2681 (not surprising, as we show here that a possible UV flare of the same luminosity as seen in NGC~4552 would have escaped detection in NGC~2681 due to its steeper surface brightness profile). 

In this paper we demonstrate that for this galaxy convincing evidence for a central BH comes instead from the modeling of the circumnuclear ionised gas kinematics derived from the FOS spectrum, properly combined with FOC photometry. In so doing, we find we can also place interesting constraints on origin of the young stellar population in the center of this galaxy. Section~\ref{sec:observations} presents the known information about NGC~2681 and the observations we obtained. Section~\ref{sec:spectroscopy} presents the spectroscopic observations. Section~\ref{sec:focanalysis} discusses the analysis of photometric HST data.  Section~\ref{sec:nucleus} derives the physical parameters for the galaxy inner nucleus, including our best estimate for upper limit to the mass of the central BH that resides in this galaxy.  Section~\ref{sec:conclusions} discusses and summarizes our results.

\section{OBSERVATIONS AND DATA\label{sec:observations}}

\subsection{Existing Data for NGC~2681\label{sec:existdata}}

NGC~2681 is a bulge-dominated S0/a galaxy at an estimated distance of 13.3 Mpc \citep{tul88}, corresponding to a linear scale of $\approx64$ pc arcsec$^{-1}$ and an absolute $B$ magnitude $M_{\rm B}\simeq-19.74$ \citep{dev91}. The value of the central magnesium index \citep[Mg$_2$=0.15;][]{bur88,jab96} is low compared to values typical of spheroids of comparable luminosity.  The IUE spectrum of this galaxy \citep{bur88} shows that the stellar population in its central region has a ultraviolet spectrum slowly decreasing towards shorter wavelengths, which is interpreted as evidence for a recent ($\approx1$ Gyr) burst of star formation \citep{bur88}.

NGC~2681 has two concentric, misaligned bars \citep{woz95} as seen in ground based images. A third inner bar is seen in this galaxy in HST/NICMOS images \citep{erw99} . This innermost bar is also seen in our higher resolution FOC/96 images. Note that NGC~2681 is viewed almost face-on and thus these structures cannot be due to projection effects (i.e. we cannot misinterpret a change of flattening in an axisymmetric structure for a change of ellipticity due to a bar).

Ground based $B-I$ color maps do not show evidence of a ring or a grand design spiral, although two barely visible, small spiral arms exist in the disk. Based on our FOC/96 images the, spiral structure pointed out by \citet{woz95} appears to extend to within the central few arcsecs of this galaxy. Galactic absorption for this galaxy is $E(\bv)\simeq0.02$ \citep{bur84,sch98}.  Ground-based \ha\ images also reveal a few \hii\ regions and strong nuclear emission \citep{woz95}. \citet{bok99} has reported detection of Centrally-concentrated emission ($r\lesssim2\farcs5$) via Pa~$\alpha$ imaging using HST/NICMOS. On the basis of their spectroscopic optical observations, \citet{ho97} classify the nucleus of this galaxy as a LINER.

\subsection{HST Observations\label{sec:hstobs}}

FOS/red detector spectra of the nucleus of NGC~2681 were obtained in 1997, using the $0\farcs21\times0\farcs21$ square aperture (0.25-PAIR). A complete log of FOS observations is given in Table~\ref{tab:fos_log}. Analysis of the FOS acquisition steps assures us that the nucleus of NGC~2681 was correctly centered in the FOS aperture (with a 0\farcs025 accuracy). This is confirmed by the fact that the central flux measured within the FOS aperture on our COSTAR-Corrected F342W FOC/96 image agrees with the flux predicted for that FOC filter from the FOS G270H spectrum. 

FOC observations of NGC~2681 include our own 1993 Pre-COSTAR and subsequent 1997 COSTAR-corrected, FOC/96 images. A detailed summary of the FOC instrumental configurations and exposure times is given in Table~\ref{tab:obs_log}.  In this work we also make use of an archival NICMOS image taken on 1997 June 7 with the F160W (space-based H-band), during one of the NIC3 ``campaigns'' \citep{bok99}. The scale of this image is 0\farcs2 pixel$^{-1}$ and the field of view (FoV) is $51\arcsec\times51\arcsec$. All of the spectra and images used in this work have been de-archived and recalibrated using the most recent version of the STScI data pipeline and the latest calibration files.

\section{FOS SPECTROSCOPY\label{sec:spectroscopy}}

\subsection{The Overall Features of the FOS Spectrum\label{sec:fosspect}}

\begin{figure}
\epsscale{0.8}
\plotone{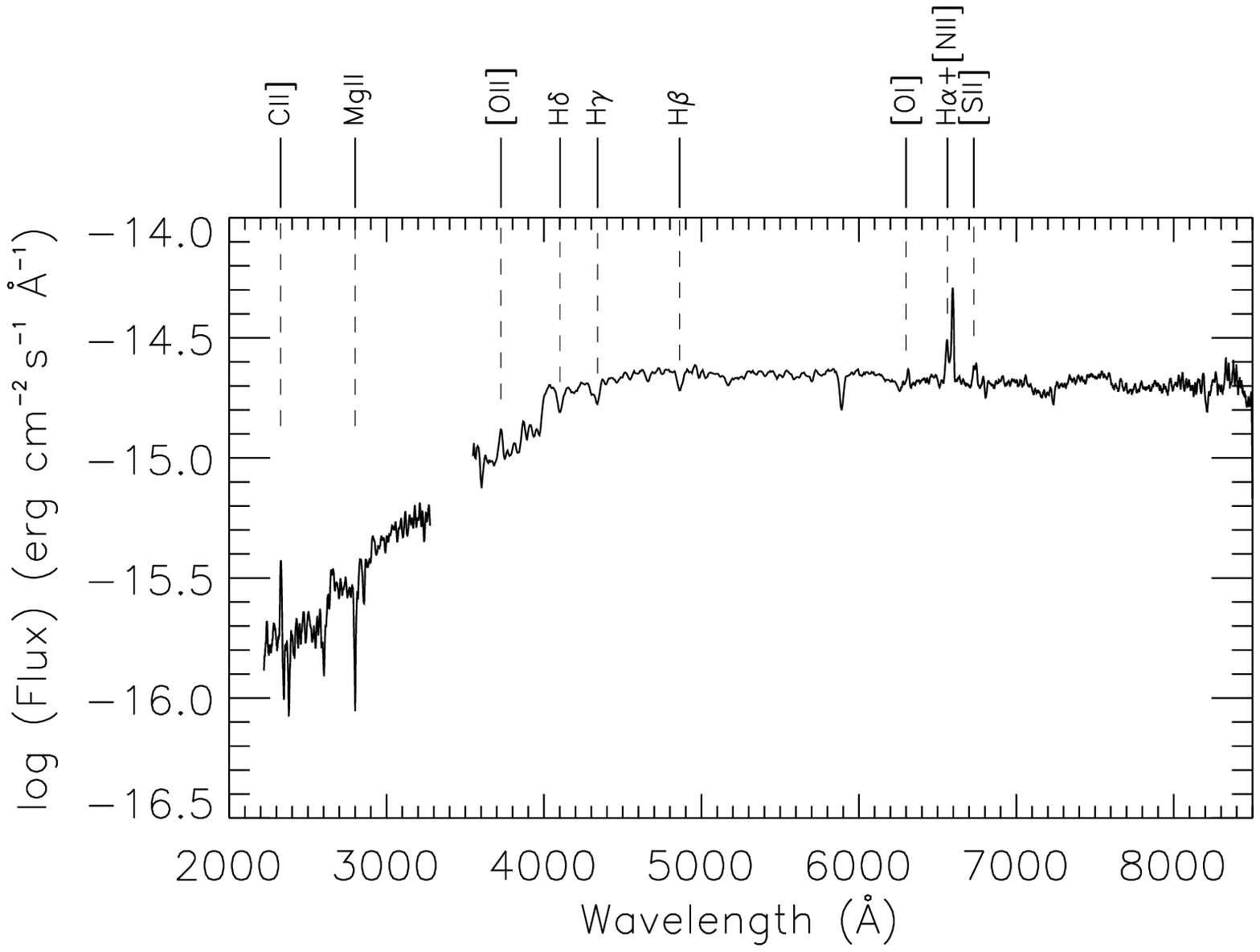}
\figcaption{Combined G270H, G650L and G780H FOS nuclear spectrum of NGC~2681 within the $0\farcs21\times0\farcs21$ aperture. All spectra have been smoothed with a Savitzky-Golay algorithm \citep{pre92}. The absorption and emission features discussed in the text are marked.\label{fig:fos_complete}}
\end{figure}

The combined FOS nuclear spectrum of NGC~2681 is presented in Figure~\ref{fig:fos_complete}.  Clearly recognizable in absorption are the Mg {II} $\lambda$2800, Mg I $\lambda$2852, and the Balmer lines $H\delta$, $H\gamma$ and $H\beta$, as well as $H\alpha$ in emission.  Strong Balmer absorption lines in the spectrum of a galaxy are, of course, indicative of a relatively young stellar component (1--2 Gyr).  This is consistent with the relative absorption line strengths of Mg II and Mg I, which are indicative of a late A-type main sequence for the stars producing this spectrum \citep{wu83}, as well as with the interpretation of IUE spectrum of this galaxy \citep{bur88}.  Also prominent are emission lines of the \cii\ $\lambda$2326 multiplet, \oii\ $\lambda$3727, \oi\ $\lambda$6300, \nii\ $\lambda\lambda$6548,6584, and \sii\ $\lambda\lambda$6716,6731.

\begin{figure}
\epsscale{0.8}
\plotone{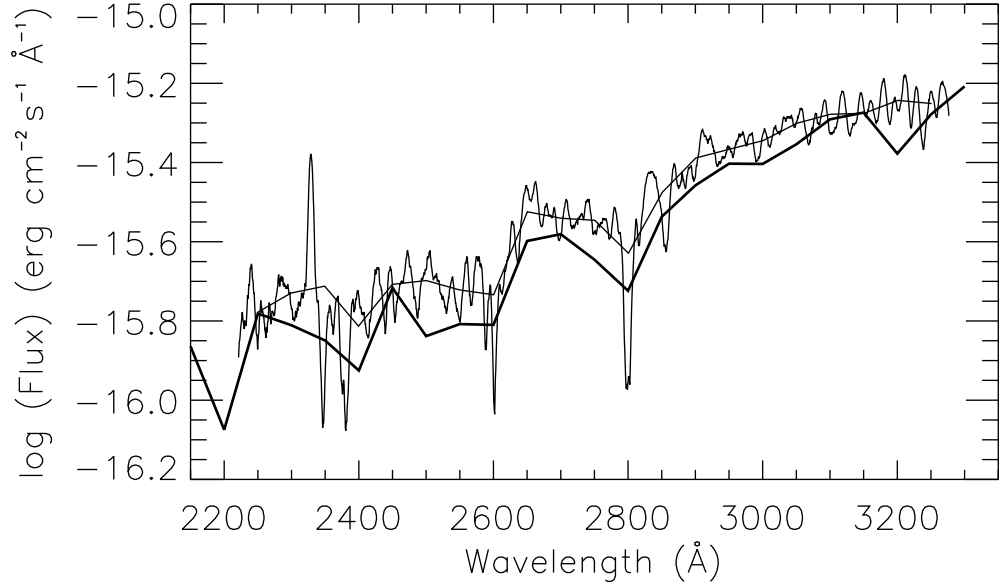}
\figcaption{Detailed comparison between the UV portion of the FOS spectrum within the $0\farcs21\times0\farcs21$ aperture (Savitzky-Golay smoothed) and IUE spectrum of NGC~2681 \citep{bur88} within the $10\arcsec\times20\arcsec$ aperture (thick line). The IUE spectrum has been rescaled by multiplying its flux by the ratio between the FOS spectrum average in the 5400--5600 \AA\ range and the V-flux within the IUE aperture quoted by \citet{bur88} for NGC~2681. An additional rebinned FOS spectrum (thin line) is superposed to the original spectrum to permit an easier comparison with the IUE energy distribution at the same resolution. Note the relative absorption line strengths of Mg II (2800 \AA) and Mg I (2852 \AA), which indicate that the main sequence stars in this part of the galaxy are late A. \label{fig:continuo}}
\end{figure}

Figure~\ref{fig:continuo} shows a comparison between the nuclear FOS spectrum of NGC~2681 and the spectrum for this galaxy within the $10\arcsec\times20\arcsec$ IUE aperture, both normalized to their respective visual fluxes. The IUE aperture covers an area $\approx4000$ times larger than the FOS aperture we used, and the FOS flux contributes $\approx4\%$ of the IUE flux. This can be verified by comparing the UV flux in the FOS and IUE spectra \citep{bur88}.  Yet, the two spectra define the same kind of relatively young stellar population within combined errors of the IUE spectrum (0.25 mag maximum error quoted by Burstein et al.) and the 5\% relative error of the FOS spectrum as quoted in the latest HST Data Handbook (Version 3).

The luminosity-weighted average radius within the FOS aperture is $R_a\approx0\farcs06$, while for the IUE aperture it is $R_a\approx5\farcs2$.  The agreement between the IUE and FOS spectra thus gives strong evidence that little, or no color gradient exists for the stellar population in the center region of NGC~2681.  As a result, we conclude that whatever redleak does exist for the UV FOC images, the lack of a color gradient justifies fitting the same two-dimensional light distribution to all the FOC images and the NICMOS image (Section~\ref{sec:focanalysis}).  Interestingly, both the lack of any perceptible dust lanes seen in the FOC images, combined with the lack of a color gradient seen in the IUE to FOS comparison, strongly suggest that the inner 0.6 kpc of NGC~2681 contains little dust.

\subsection{The Emission Lines\label{sec:emission}}

\begin{figure}
\epsscale{0.8}
\plotone{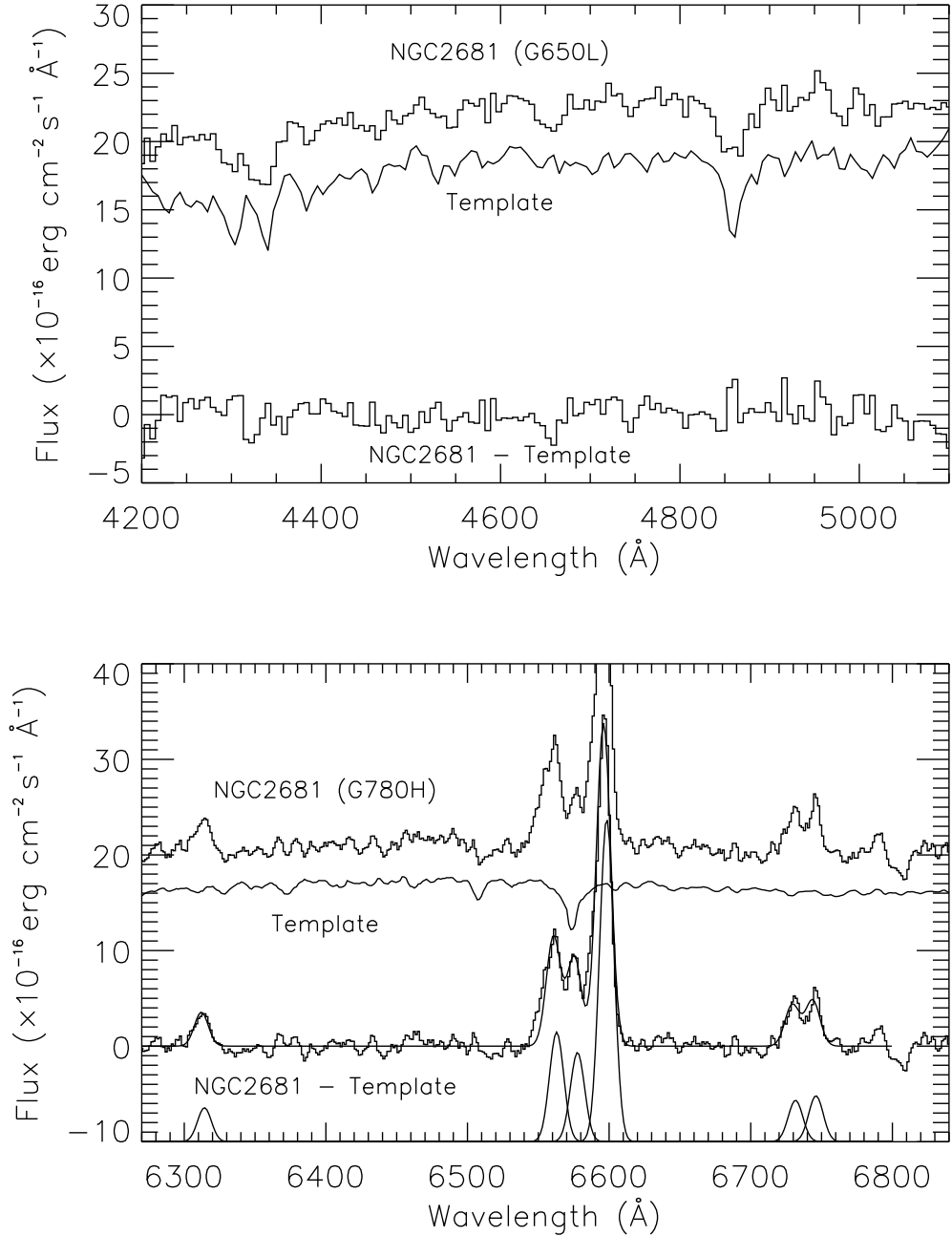}
\figcaption{{\em Upper panel:} Template--subtracted nuclear FOS G650L spectrum of NGC~2681. The upper, middle and lower plots show the original spectrum, the vertically shifted template constructed as discussed in the text, and the difference between the above two spectra, respectively. Only emission at H$\beta$ is apparent in this relatively low signal-to-noise spectrum. {\em Lower panel:} Same as in the upper panel for the grating G780H. Superposed to the lower, template-subtracted spectrum is our resulting model including the \oi, \ha, \nii\ and \sii\ emission lines. The individual Gaussian emission components of the model are shown at the bottom of the panel. \label{fig:modeling}}
\end{figure}

The emission lines in our spectra have been analyzed by a ``template subtraction'' procedure as in \citet{ho97}. Starting from the high S/N spectra of \citet{ho95}, we have used a minimization algorithm to construct an emission-line-free template for their NGC~2681 2''$\times$4'' aperture spectrum. The best match was obtained using a combination of their spectra of NGC~205 and NGC~4339. We use this template to subtract the underlying stellar contribution from the lower S/N 0\farcs2$\times$0\farcs2 FOS spectrum. The upper panel of Figure~\ref{fig:modeling} illustrates the template subtraction from the G650L spectrum (lacking any reliable emission), while the lower panel shows both the template subtraction and the subsequent fit to the prominent emission lines of the template-subtracted FOS G780H spectrum.

The emission lines in the continuum-subtracted optical FOS spectrum have been modeled using the Levenberg-Marquardt algorithm \citep{pre92} to fit a sum of Gaussian components to the observed spectrum. We adopt the minimum set of free parameters able to generate a satisfactory fit by assuming that all emission lines can be fit by a Gaussian profile with the same redshift and the same velocity width within a given grating. The derived parameters for the emission lines are given in Table~\ref{tab:line_modeling}. The $1\sigma$ formal errors of the line parameters are estimated from the covariance matrix. The \cii\ emission line has also been fit in the G270H spectrum (Table~\ref{tab:line_modeling}) by assuming a linear template under the emission feature.

The integrated \ha$+$\nii\ flux in our $0\farcs21\times0\farcs21$ aperture is only a factor $\approx1.9$ smaller than the flux measured by \citet{ho97} within a $2\arcsec\times4\arcsec$. Given that the ratio of the two apertures is a factor $\approx180$, this small difference in the measured flux shows that the gaseous emission is very centrally concentrated and, in particular, comes mainly from a region not exceeding in size the FOS aperture.

\subsection{Emission Line Diagnostics\label{sec:emissdiag}}

\begin{figure}
\epsscale{1}
\plotone{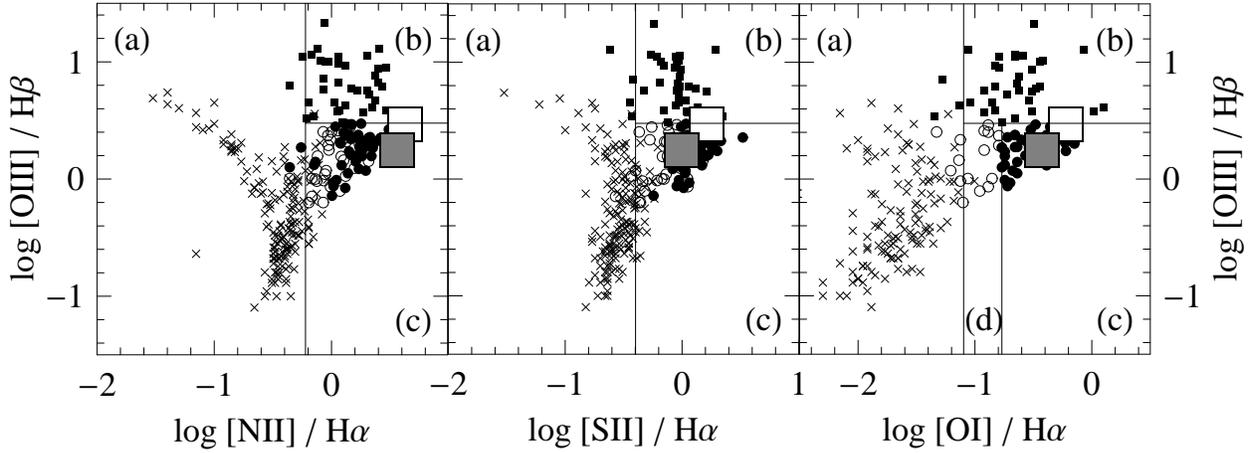}
\figcaption{Location of the NGC~2681 nucleus (large gray square) on the diagnostic diagrams of \citet{ho97}, as derived from emission line ratios measured from the FOS spectra. The observational errors are comparable with the size of the smaller symbols. The large open square marks the position of the NGC~4552 nucleus, as determined by \citet{cap99}. The other symbols represent the nuclei included in the Ho et al. sample (crosses = \hii\ nuclei, filled squares = Seyfert nuclei, filled circles = LINERs, open circles = transition objects). The vertical and horizontal lines delineate the boundary adopted by Ho et al.\ between (a) \hii\ nuclei, (b) Seyfert galaxies, (c) LINERs and (d) Transition objects.\label{fig:diagnostics}}
\end{figure}

Figure~\ref{fig:diagnostics} compares the emission line ratios obtained from our HST spectra of the nuclei of NGC~2681 and NGC~4552 \citep{cap99}, to those for Seyfert galaxies, LINERs, and \hii\ galaxies from \citet{ho97}. Note that the \oiii\ and \hb\ fluxes were not measured from our G650L FOS spectrum, as these lines have low $S/N$ in our data. Rather, for these emission lines, we resort to the \citet{ho97} ground-based values. This choice cannot affect the physical classification of NGC~2681 nucleus as most of the flux observed by Ho et al. comes from a region within our HST aperture and---what is more---the basic discrimination on the AGN/non-AGN nature of the nucleus comes from our own (FOS) emission line ratios.

As can be seen, the nucleus of NGC~2681 is clearly classified as a LINER from our FOS spectroscopy, in accordance with the ground based classification of \citet{ho97}. We also note with interest that, despite the structural differences of the central regions of the two galaxies, the location of the innermost source of NGC~2681 is surprisingly close to that of the variable spike in NGC~4552 in these spectroscopic diagnostic diagrams.

\subsection{Constraints on the Size of the Emitting Region in NGC~2681\label{sec:emiss-size}}

A rough estimate of the size of the line-emitting region can be obtained from the density as set by the forbidden lines, and by the \ha\ luminosity \citep{ost89}:
\begin{equation}
    V\simeq\frac{L({\rm H}\alpha)}{f\ n_{\rm e}^2\ \alpha_{{\rm
     H}\alpha}^{\rm eff}\ h\ \nu_{{\rm H}\alpha}},
\end{equation}
where $V$ is the volume occupied by the emitting gas, $L({\rm H}\alpha)$ is the total luminosity of the H$\alpha$ line, $f$ is the volume filling factor, $n_{\rm e}$ is the electron density, $\alpha_{{\rm H}\alpha}^{\rm eff}=1.17\times10^{-13}$ cm$^3$ s$^{-1}$, $h$ is the Planck constant and $\nu_{{\rm H}\alpha}$ the frequency of the H$\alpha$ line.

Adopting $n_{\rm e}\simeq1860$ from \citet{ho97}, a filling factor $f=1$, a distance $d=13.3$ Mpc \citep{tul88}, and spherical geometry, we obtain $L({\rm H}\alpha)\approx2\times10^{38}$ erg s$^{-1}$ and a size of the emitting region $r\approx1.2$ pc ($\approx0\farcs02$). This estimate is within the upper limit imposed by the size of our FOS aperture ($\approx6$ pc). The latter size would correspond to a much lower filling factor of $\approx5\times10^{-3}$.

\section{FOC AND NICMOS PHOTOMETRY\label{sec:focanalysis}}

Basic ingredients of the dynamical modeling we present later are the availability of a high resolution surface brightness profile of the galaxy, as well as observational support to the adopted hypothesis of a constant stellar mass-to-light (M/L) ratio. We show here that the data provided by the FOC and NICMOS images do assure both pieces of needed information.

Ideally, in order to obtain reliable surface brightness profiles and color gradients, one should correct the FOC/96 images for nonlinearity effects in the FOC \citep{not96}, deconvolve all images with the PSF and, finally, derive the photometric profiles and the corresponding color gradients from the corrected frames. However, in our case this process is numerically very unstable, owing both to the very low signal-to-noise ratio of the FOC images and the approximations introduced by the nonlinearity correction which are greatly amplified by the deconvolution process.

\begin{figure}
\epsscale{0.8}
\plotone{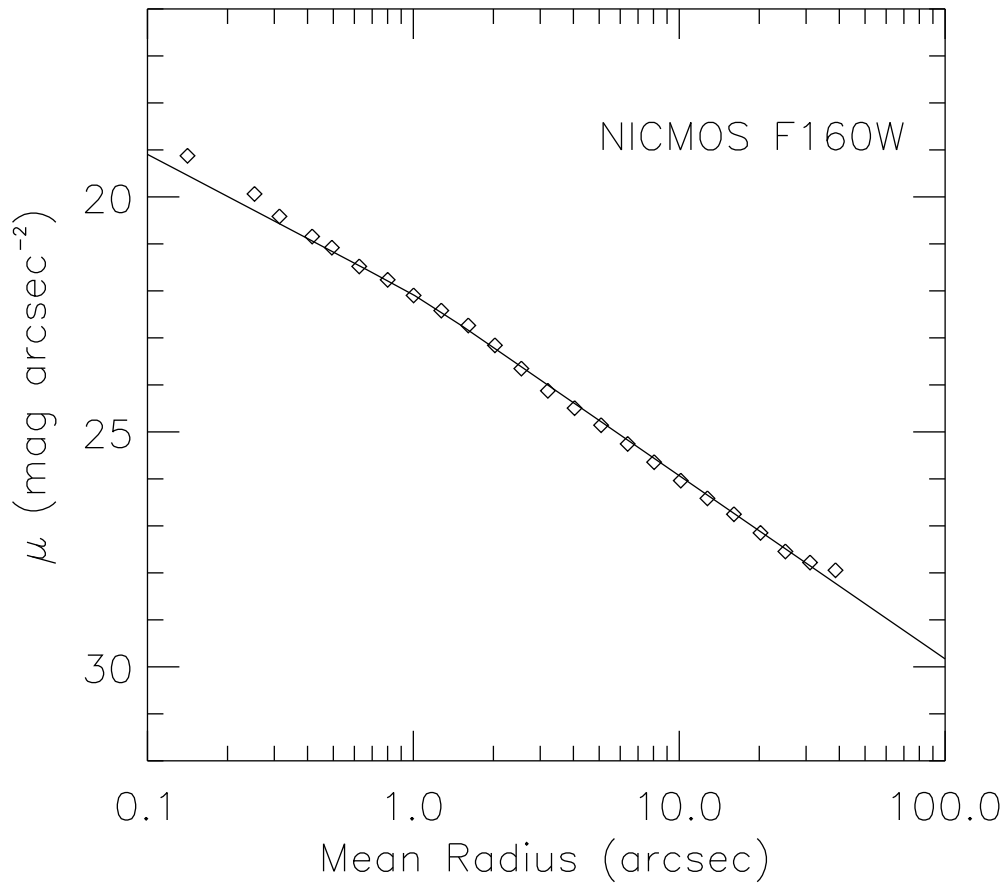}
\figcaption{The surface brightness profile of NGC~2681 on the STMAG photometric system, measured from the F160W NICMOS image. Diamonds represent the observed values, while the solid line represents the double power-law profile of the best fitting photometric model, determined as described in the text. Note that exactly the same double power-law profile has been used to model the FOC photometry in Figure~\ref{fig:foc}.\label{fig:nicmos}}
\end{figure}

\begin{figure}
\epsscale{0.6}
\plotone{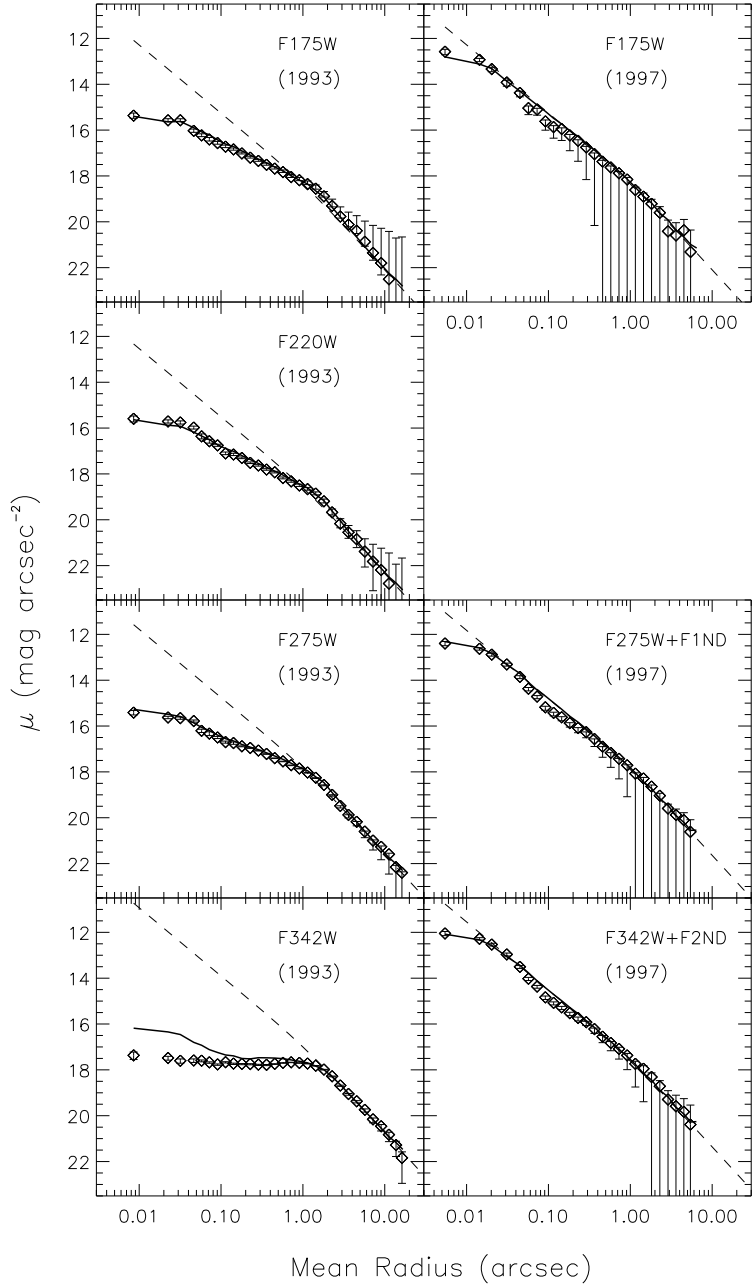}
\figcaption{STMAG photometric profiles in the FOC/96 passbands. Dashed lines always represent (apart from a scale factor) the unique double power-law model, fitting simultaneously all the HST profiles. Such a unique model translates into different expected profiles (solid lines) once the instrumental effects affecting each different FOC image are applied. Diamonds represent the measured profiles after subtracting the most appropriate dark level. Error bars represent the allowed range of surface brightness at each measured radius, given the possible range of background count rate and measured random errors for each image. The fact that in most cases diamonds lie close to the upper bar limit comes from the fact that the dark-rate was generally close to the nominal (lowest) value. The same double power-law profile also reproduces the NICMOS data in Figure~\ref{fig:nicmos}. The highly nonlinear pixels within $\approx1\arcsec$ in the F342W 1993 image cannot be reproduced by the model.\label{fig:foc}}
\end{figure}

The similarity of the available FOS and IUE spectroscopy (Section~\ref{sec:spectroscopy}) strongly suggests a lack of detectable color gradient between $R\approx0\farcs06$ and $R\approx5\farcs2$. Moreover, the good agreement between the $U$ (FOC/F342W) and $H$ (NICMOS/F160W) profiles in the radial range 1\arcsec--10\arcsec---not affected by dark subtraction, nonlinearity and redleak effects---confirms the absence of any significant stellar population gradient in the central regions of this galaxy (see Figures~\ref{fig:nicmos},~\ref{fig:foc}). For these reasons we feel permitted to construct a {\em unique} bi-dimensional photometric model of the center of NGC~2681.

The geometric properties of the model are fixed by imposing a match with the observed isophotes as obtained with the \verb#ellipse# IRAF task. The photometric profile is derived as a function of the free parameters given below. This model is then convolved with the proper PSF---a normalized mean of both UV and red-leaked PSFs constructed by means of Tiny Tim \citep[version 4.4]{kri92} and weighted in accordance to a properly chosen energy distribution---taking correctly into account of nonlinearity effects \citep[for FOC images alone; see Appendix of][]{cap99}. The resulting model is then simultaneously fit by means of an iterative procedure to all the original images as given by the STScI pipeline (with the exclusion of the highly nonlinear pixels with $R<1\arcsec$ in the F342W 1993 image).

\begin{figure}
\epsscale{0.8}
\plotone{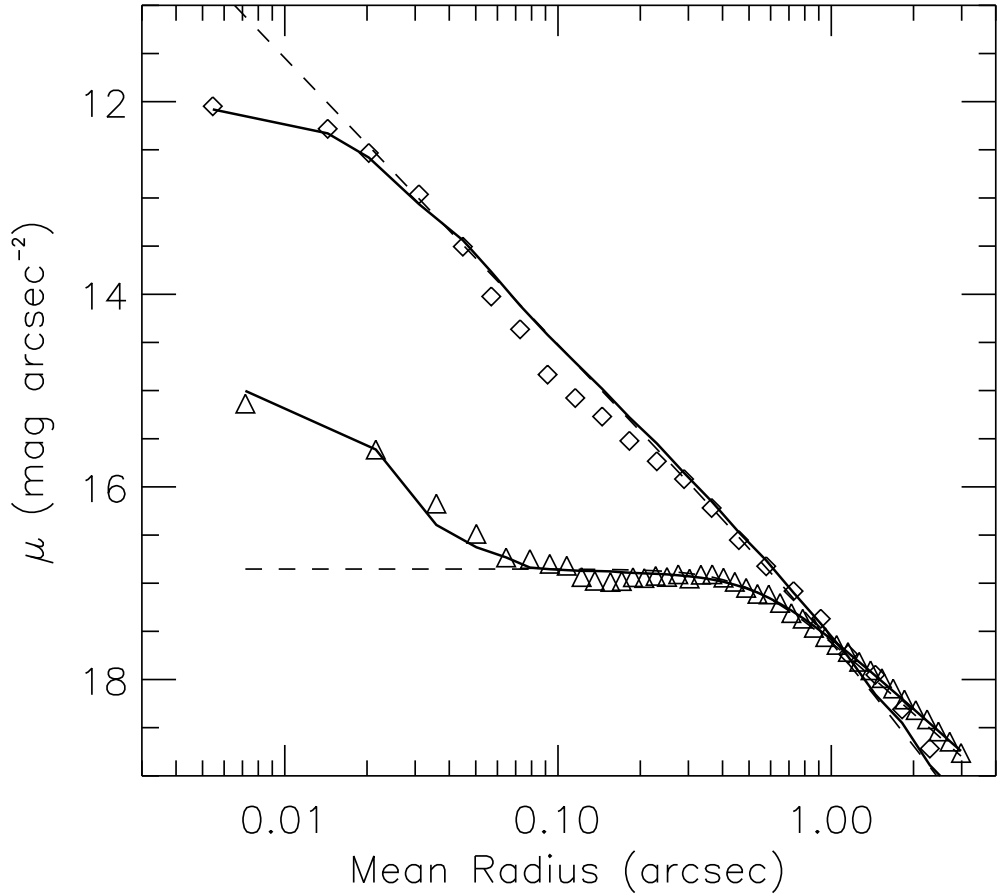}
\figcaption{Comparison between the 1996 FOC/96 COSTAR-corrected photometric profiles of NGC~4552 (triangles) in the F342W band and NGC~2681 (diamonds) in the F342W+F2ND band from our 1997 FOC observation. For each profile (shown in the STMAG photometric system) open symbols represent the observed points, the dashed line is the double power-law profile of the best fitting photometric model of the galaxy and, finally, the solid line represents the model after the inclusion of FOC instrumental effects. A point-like source has been added to the NGC~4552 model to reproduce the observations (see \citet{cap99} for details).\label{fig:profiles_comparision}}
\end{figure}

A very good fit is achieved also for the far--UV passbands, (Figure~\ref{fig:foc}), consistent with the lack of color gradients. However, only $\approx$50\% of the detected flux in the three red-leak-affected FOC filters (F175W, F220W and F275W) is genuinely UV light (i.e., falls within the filter FWHM, as one can verify by using their transmission curves and our own FOS spectrum), thus weakening the importance of this latter outcome, when compared to the evidence quoted above.  The perfect match obtained with both the 1993 and 1997 FOC observations also assures that the nucleus of this galaxy has not significantly varied in surface brightness between the two epochs. However, it is worth mentioning that the FOC/F342W profiles of NGC~2681 and NGC~4552 (Figure~\ref{fig:profiles_comparision}) show that the surface brightness of NGC~2681 close to the center is higher by $\approx4$ mag arcsec$^{-2}$ than that of NGC~4552 for $R\approx0\farcs03$, before the variable spike starts dominating its luminosity. As a consequence the central UV surface brightness of NGC~2681 is so high that a UV flare of the luminosity seen in NGC~4552 would have not been detected with observations of the kind discussed here, leaving the NGC~2681 nucleus apparently quiescent for the observer.

We find that a unique double power-law \citep{lau95}
\begin{equation}
    \label{eq:nuker} I(R) = I_{\rm b}\;
    2^{\frac{\beta-\gamma}{\alpha}} \left(\frac{R}{R_{\rm
    b}}\right)^{-\gamma} \left[1 + \left(\frac{R}{R_{\rm
    b}}\right)^{\alpha}\right] ^{-\frac{\beta-\gamma}{\alpha}},
\end{equation}
with parameters $2.0\lesssim\alpha$, $\beta=1.56\pm0.02$, $\gamma=1.19\pm0.02$, $R_b=1\farcs1\pm0\farcs1$ can reproduce all of the eight profiles (see Figures~\ref{fig:nicmos} and \ref{fig:foc}) and also satisfy the constraint that the FOC dark countrate (which is also a free parameter for the FOC frames) remains within plausible ranges \citep{not96}. Here $\gamma$ measures the steepness of the inner profile ($I(R) \propto R^{-\gamma}$ for $R \ll R_{\rm b}$), $\beta$ is the steepness of the outer profile ($I(R) \propto R^{-\beta}$ for $R \gg R_{\rm b}$), $\alpha$ is related to the sharpness of the transition and $I_{\rm b}$ is the scale factor for luminosity. In the profiles a radius of 0.38 pixel---corresponding to the average radius within the central pixel---has been assigned to the central value of the surface brightness. Due to the small FWHM and the good sampling of the PSF in the UV our FOC/96 non-aberrated UV images provide information down to the central $\approx0.3$ pc of NGC~2681.

\begin{figure}
\epsscale{0.8}
\plotone{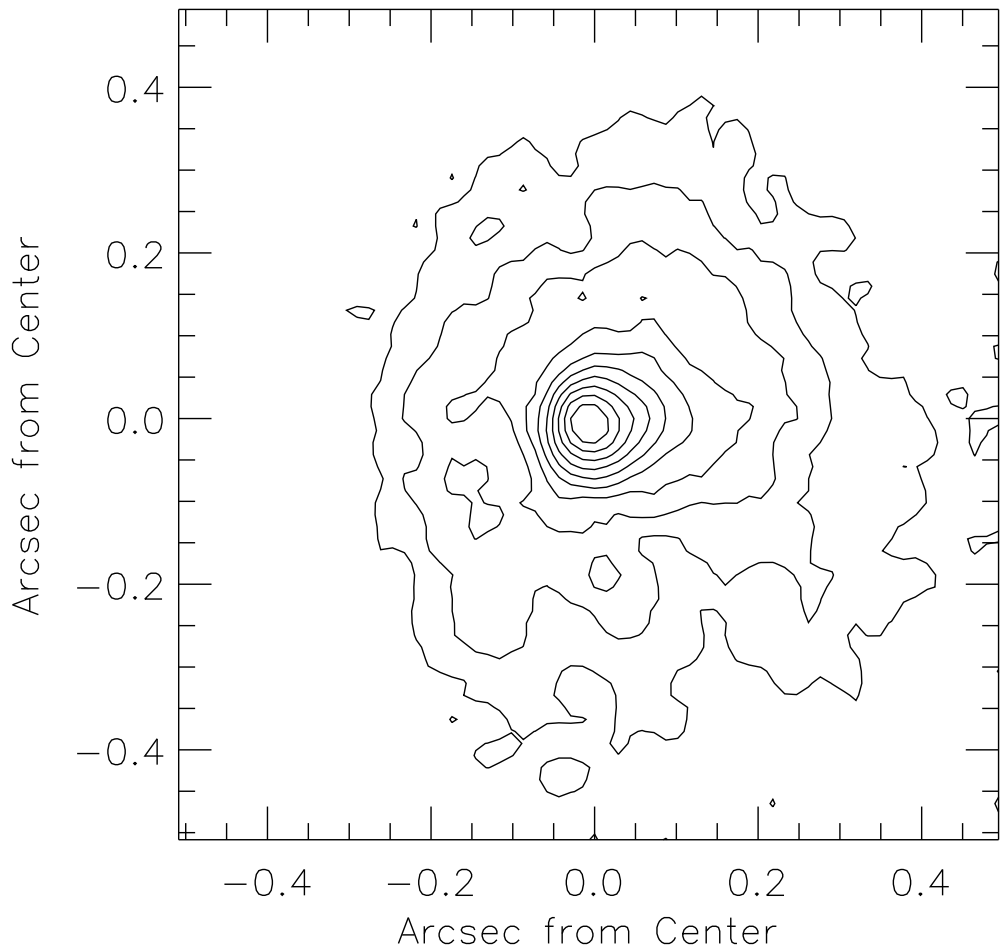}
\figcaption{Contour image of the central arcsec in the 1997 FOC/96 non-aberrated image of NGC~2681. This image has been obtained by accurately aligning the three F175W, F275W+F1ND and F342W+F2ND FOC/96 images by cross-correlating the frames and then coadding them to improve the signal-to-noise. The resulting image has been finally boxcar-smoothed over $3\times3$ pixels ($0\farcs04\times0\farcs04$). The contours are logarithmically spaced. Note the clear asymmetry of the nucleus and the irregular appearance of the surrounding regions. \label{fig:contours}}
\end{figure}

The profiles in Figures~\ref{fig:nicmos} and \ref{fig:foc} represent the surface brightness in mag arcsec$^{-2}$ in the STMAG magnitude system. The magnitude zeropoints---computed using the SYNPHOT package within IRAF/STSDAS---take into account the 1.27$\times$ higher sensitivity of the FOC zoomed mode, relative to the non-zoomed observations \citep{cap99,not96}.  It is evident from the profiles from the 1997 FOC images (Figure~\ref{fig:foc}, right panels) that there is a change of slope of the observed profile around $R\approx0\farcs1$ which is {\em not} reproduced by the model. This radius corresponds to the distance where the nucleus becomes more asymmetric (see Figure~\ref{fig:contours}). This, in turn, suggests that some substructure could be present. However, no difference in color or spectral energy distribution is observed for this central feature, thus justifying our choice of resorting to a single component density profile when modeling the dynamics of the central regions.

\section{DYNAMICAL CONSTRAINTS ON THE NUCLEAR MASS DISTRIBUTION\label{sec:nucleus}}

In order to construct a dynamical model of the innermost regions we need to extract a reliable intrinsic density profile, starting from the observed photometry. To this aim the analytic double power-law photometric profile derived in Section~\ref{sec:focanalysis} is deprojected by means of an Abel inversion \citep{bin87}, under the assumption of spherical symmetry. The profile is converted to the V-band using the color $\mu_{342}-\mu_V\simeq1.0$ mag measured from the nuclear FOS spectrum of Figure~\ref{fig:fos_complete} (the STMAG photometric system has the same zero point as that of the V magnitude band) and applying a foreground extinction $A_V=0.07$ \citep{bur84,sch98}.

\begin{figure}
\epsscale{0.8}
\plotone{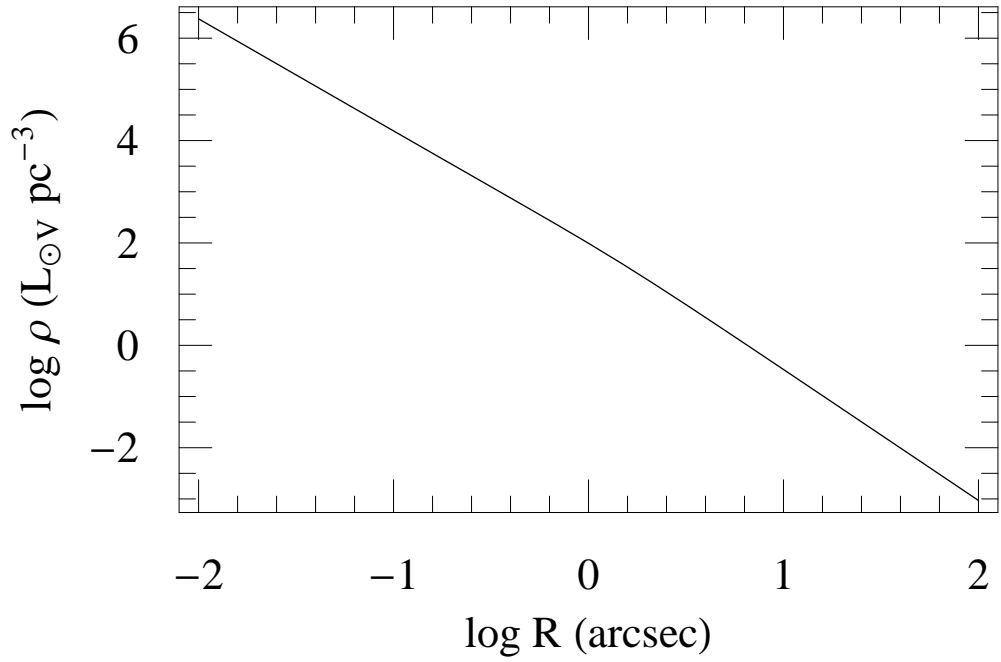}
\figcaption{Deprojected density profile of NGC~2681. The profile has been obtained by Abel inversion of the best-fitting analytic double power-law profile. The F342W+F2ND band calibrated profile has been first converted to the V band by subtracting the value $\mu_{342}-\mu_V\simeq1.0$ mag, measured from the FOS spectrum. \label{fig:density}}
\end{figure}

The deprojected luminosity density profile in units of $L_{\sun V}$ pc$^{-3}$ is presented in Figure~\ref{fig:density}. The derived density in the central pixel ($\approx0\farcs01$) corresponds to $\rho_0\gtrsim2.4\times10^6$ $L_{\sun V}$ pc$^{-3}$. This is quite a high value for the central luminosity density. For comparison, for a sample of 41 spirals analyzed by \citet{car98}, the highest value of the density at 10 pc from the nucleus is a factor $\approx2$ lower than the corresponding value we derive for NGC~2681 ($\rho_{10 \rm pc}\approx6\times10^3$ $L_{\sun V}$ pc$^{-3}$).

Velocity dispersion and rotation curve profiles are not available for NGC~2681. Moreover, as this galaxy contains three concentric bars \citep{erw99}, developing a fully three-dimensional model of its mass distribution is problematic, at best. However, experience has shown that, while detailed observations and general dynamical models are needed to establish the presence of possible  BHs in galaxies, a simple isotropic dynamical model gives reasonable order of magnitude estimates of the BH parameters. A classic example is the $\approx5\times10^9$ \msun\ BH in M87 inferred by \citet{sar78} using spherical isotropic models, an estimate reasonably close to the most recent values ($\approx3\times10^9$ \msun) determined by \citet{har94} and \citet{mac97} using HST spectra of the gaseous disk surrounding the nucleus of M87.

A detailed analysis of spherical models with cuspy power-law inner surface density profiles $I(R)\propto R^{-\gamma}$ is available from \citet{deh93}, \citet{tre94} and \citet{zha96,zha97}. The dynamics of isotropic power-law cuspy nuclei (without BH) are dramatically different, depending on whether $0<\gamma<1$ or $1<\gamma<2$ ($\gamma$ being the inner slope of the power-law). For the lower $\gamma$ case, termed ``weak cusp'' by \citet{tre94}, the mean-squared line-of-sight velocity dispersion $\sigma^2_p$ approaches zero near the center. For the other case, termed the ``strong cusp'' regime, the relation $\sigma^2_p\propto R^{1-\gamma}$ diverges near the center (finite resolution will actually transform these mathematical singularities to finite observable values if $\gamma<1.5$).

In either case, if a BH is present $\sigma^2_p\propto R^{-1}$ near the center. Thus it is clear that, while with $0<\gamma<1$ an increase of $\sigma^2_p$ in the very inner regions ($R\ll R_b$) is already an indication of a possible BH, this is not the case when $1<\gamma<2$, as is such for NGC~2681 ($\gamma\simeq1.19$). A BH becomes finally undetectable for $\gamma\rightarrow2$, when its presence induces the same trend as a simple increase of the M/L stellar ratio.

\begin{figure}
\epsscale{0.8}
\plotone{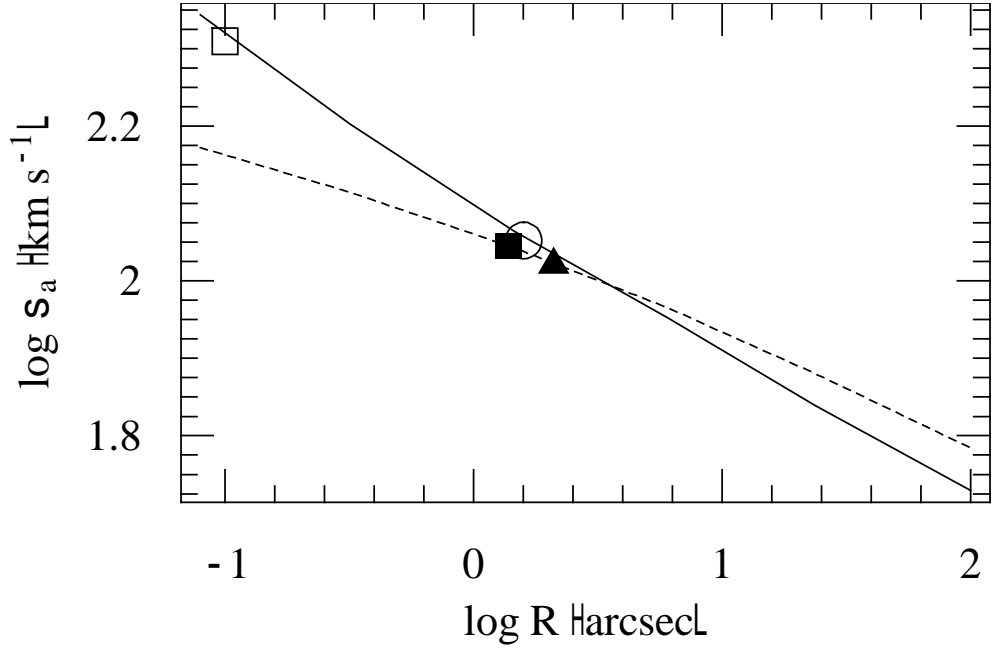}
\figcaption{Aperture velocity dispersion ($\sigma_{\rm a}$) profile (dotted line) computed by deprojecting the observed double power-law photometric profile and by solving the hydrostatic equilibrium equation under the assumption of spherical symmetry and isotropy of the velocity-dispersion tensor. With $\Upsilon=1.0~\msun/\lsun_{V}$ this profile agrees with the stellar dispersion measured by \citet[filled square]{dal91} and by \citet[filled triangle]{sch83}. The open circle and open square indicate the gaseous dispersion measurement by \citet{ho97} and our own FOS value, respectively. The solid line represents the same isotropic model with the addition of a $6\times10^7$ \msun\ central BH and $\Upsilon=0.7$. Since a double power-law brightness profile with $\gamma\geq1$ and a central BH would generate an infinite observed dispersion, in this computation we have assumed that the stellar density profile becomes constant inside the FOC/96 resolution limit (see text for details). \label{fig:dispersion}}
\end{figure}

Figure~\ref{fig:dispersion} shows the kinematical data available for this galaxy: ground based measurements of the {\em stellar} central velocity dispersions of \citet[$\sigma=111\pm16$ \kms\ within a $1\farcs5\times4\arcsec$ aperture]{dal91} and \citet[$\sigma=105\pm9$ \kms\ within a 4\farcs2 diameter aperture]{sch83}, and the {\em gas} velocity dispersion of \citet[$\sigma=113\pm5$ \kms\ within a $4\arcsec\times2\arcsec$ aperture]{ho97}. In Figure~\ref{fig:dispersion} we also plot our own {\em gas} dispersion FOS measurement of $\sigma=204\pm5$ \kms\ within the $0\farcs21\times0\farcs21$ aperture. The central $\sigma$ is found to increase by a factor of $\approx2$ when narrowing the aperture from $R\approx1\farcs5$ (ground) to $R\approx0\farcs1$ (HST/FOS).

To investigate this further, we have computed a simple spherical isotropic model for the stellar kinematics, starting from the double power-law photometric profile $I(R)$ determined above, and assuming a constant stellar mass-to-light ratio $\Upsilon\equiv\msun/\lsun_{V}$. With these assumptions the projected stellar velocity dispersion at a distance $R$ from the center can be predicted from the relation \citep[e.~g.][]{tre94}:
\begin{equation}
\label{eq:sigma_p}
\sigma_{\rm p}^2(R) I(R)=\frac{2 G}{\Upsilon} \int_{R}^{\infty}
    \frac{\rho(r) M(r)}{r^2} (r^2-R^2)^{1/2}\; dr.
\end{equation}
The stellar density $\rho(r)$ and the mass within a given radius $M(r)$ have been obtained from the deprojected double power-law profile. $R$ is the radial distance on the plane of the sky, while $r$ is the radial coordinate measured from the center of the galaxy. In the case a BH lies at the center of the galaxy we have to substitute the enclosed mass $M(r)$ in Equation~(\ref{eq:sigma_p}) with the expression $[M(r)+M_\bullet]$, where $M_\bullet$ is the mass of the BH.

In order to compare the predictions with the observations we calculate the velocity dispersion $\sigma_{\rm a}(R)$ as measured through a circular aperture of radius $R$ {\em centered on the galaxy nucleus} (seeing and PSF effects are not modeled here, since they are smaller than the apertures used):
\begin{equation}
\label{eq:sigma_a}
\sigma_{\rm a}^2(R)=\frac{\int_0^R \sigma_{\rm p}^2(R') I(R') R'\;
dR'}{\int_0^R I(R') R'\; dR'}.
\end{equation}

We evaluate this equation for the two cases with and without a central BH. In the former case we impose that below $R=0\farcs005$ (the central resolution of our photometric {\em model}, corresponding to 0.38 FOC/96 pixels) the stellar density profile flattens, becoming constant (an extreme assumption, which maximizes the possible BH mass). We do this in order to avoid the integral at the numerator in Equation~(\ref{eq:sigma_a}) diverging. To compare the models with the data we further assume that the gas behaves as a ``hot'' system, being in distinct clumps in isotropic motion, with a density proportional to the stellar density. This is appropriate if the gas is produced by the stars themselves, a not unreasonable assumption.

Figure~\ref{fig:dispersion} shows the results of our modelling with the inclusion of a central BH of mass $M_\bullet=6\times10^7$ \msun\ (solid line) and without it (dotted line). It is apparent that a model without a BH cannot explain the observations. However, we stress that, due to the above assumption on the flattening of the central density profile, the value $M_\bullet\lesssim6\times10^7$ \msun\ represents a strict {\em upper limit} of the mass of a possible central BH in this galaxy. In this respect, one should be aware that only measurement of stellar kinematics using long slit STIS spectra will permit us to remove the present degeneracy in the modelling of NGC~2681.

We also note that the ground-based data constrain $\Upsilon$ to be 0.7 and 1.0 for the two above cases, i.e. with and without BH, respectively. A value $\Upsilon\approx1$ is consistent with simple stellar population models of solar metallicity and ages 1-2 Gyr, for both Salpeter and Scalo IMF \citep{mar98}.

The upper limit for the BH mass determined in this paper ($M_\bullet\lesssim6\times10^7$ \msun) is consistent with the recent very tight correlation between the central velocity dispersion $\sigma_c$ and the BH mass as given by \citet{mer00}. The central dispersion $\sigma_c\approx111$ \kms\ (corrected to $r_e/8$) we observe for this galaxy, corresponds in fact to an expected BH mass of $M_\bullet\approx8\times10^6$ \msun.

We remark that in the above BH mass estimate we have modeled the gas as a ``hot'' system of point masses. This appears justified from the fact that within an $R\approx1\farcs5$ aperture the dispersion observed from the gas and the stars is the same (see Figure~\ref{fig:dispersion}). The gas motion however may become more ordered closer to the nucleus. As an extreme case, one could even assume that, at HST resolution, the gas lies in a thin disk in circular motion in the combined potential of the BH and the stellar potential.

To test this possibility we have tried to explain the observed emission line profile as produced by a thin gaseous disk using the techniques we have developed in \citet{ber98}, and subsequently integrating the model spectrum within the FOS aperture. Using different BH masses and various disk inclinations, we have verified that, under the thin disk hypothesis, it becomes impossible to produce an emission line having the width $\sigma\simeq204$ \kms\ observed with FOS without producing a double peaked line profile. Intrinsic dispersion in the gas is thus important and the disk cannot be thin. This assures that the adopted spherical assumption for the geometry gives the correct order of magnitude estimate for the BH mass.

\section{CONCLUSIONS\label{sec:conclusions}}

Our analysis of the light distribution in the inner 0.6 kpc of NGC~2681 shows that a double power law light profile fits its light distribution both in the NICMOS images and in our set of FOC/96 UV images. The good fit of the multiband data obtained with the {\em same} light distribution, coupled to the close similarity of our FOS spectrum to the IUE spectrum taken with a much larger aperture, argues against the presence of sizable color gradients in the center regions of NGC~2681.  A remarkable consequence of the steep, innermost profile of NGC 2681 is that it translates---when observed in the near-UV---into a central surface brightness some 4 mag arcsec$^{-2}$ higher than that of the recently identified UV-variable spike of NGC~4552 \citep{cap99}. This implies that a UV-bright flare of comparable luminosity as that observed at the center of NGC~4552 would escape detection in the nucleus of NGC~2681.

The FOS spectrum of the central region reveals the presence of broad emission lines. Comparison to ground based spectra shows that the gaseous emission is centrally concentrated, most of it coming from a region inside the central $\approx6$~pc of NGC~2681. On the basis of the observed ratio of the emission lines, the nucleus of NGC~2681 falls in the region of LINERs. This suggests a slow, steady feeding of a central BH.

The presence of the BH is more strongly indicated by the increase of a factor $\approx2$ of the gas velocity dispersion, when narrowing the aperture from $R\approx1\farcs5$ to $R\approx0\farcs1$. Such an increase cannot be accounted for by the stellar potential alone, as derived by deprojecting the observed surface brightness profile. We also find that the gas in the inner regions of NGC~2681 is not supported by rotation, as in the case of a Keplerian disk, but mainly by random motions. Assuming that the gas can be modeled as a ``hot'' system, we derive a firm upper limit to the BH mass of $M_\bullet\lesssim6\times10^7$ \msun\ from an isotropic dynamical model. Such an estimate is consistent with the $M_\bullet-\sigma$ relation of \citet{fer00} and \citet{geb00}.

In this respect, NGC~2681, with a central velocity dispersion of 111 \kms\ (corrected to $r_e/8$) and a predicted BH mass $M_\bullet\approx8\times10^6$ \msun, must be regarded as an object of particular interest as it falls in the still poorly populated low-$\sigma$ region of the tight $M_\bullet-\sigma$ relation. One should be aware, however, that only the knowledge of the {\em radial dependence} of the velocity dispersion (and not its central value alone, as is our current knowledge for NGC~2681) assures the possibility of pinpointing the BH mass in NGC~2681 instead of establishing its upper limit, as we have done here.  To acquire such data requires HST/STIS {\em slit} observations.

Our dynamical models also indicate a stellar mass to light ratio for the central population of 0.7-1, which is appropriate for a 1-2 Gyr old, solar metallicity stellar population. This is fully consistent with previous IUE and optical spectroscopy (namely, the measured Mg$_2$ line strength index) indicating that the population of this galaxy is dominated by {\em coeval} stars formed in a fairly-well defined starburst $\approx1$ Gyr ago \citep{bur88}.

Finally we remark that a HI survey of NGC~2681 and its associated galaxies would be highly desirable.  Such a survey might reveal evidence that NGC~2681 underwent a tidal interaction with another galaxy in the recent past, lending support to the hypothesis of a major merging event affecting what we see in the center of this galaxy today.  On the one hand, NGC~2681 belongs to a galaxy group, an environment in which mergers are particularly favored to occur. On the other hand, NGC~2681 does not show clear signatures of the recent accretion of a gas-rich dwarf, such as morphological distortions or duplicity of the nucleus.  While the presence of a uniform, dustless, young stellar population in its central 0.6~kpc is puzzling, perhaps we can view NGC~2681 as the older brother of M82, seen face-on, after the star burst has run its course.

\acknowledgements

MC is grateful to the European Southern Observatory for its kind hospitality in the period during which much of the HST data analysis has been completed. DB acknowledges support by NASA/STscI through grants GO-03728.01-91A and GO-06309.01-94A.



\clearpage

\clearpage

\begin{deluxetable}{ccccc}
\tablecaption{FOS Observation Log.\label{tab:fos_log}} \tablehead{
\colhead{Grating} & \colhead{$\lambda$ range} & \colhead{Spectral
Resolution} & \colhead{Date} & \colhead{Exp.} \\
& \colhead{(\AA)} & \colhead{(FWHM \AA)} & & \colhead{(s)} }
\startdata
G270H & 2222--3277 & 1.89 & 1997 Feb 2 & 2600 \\
G650L & 3540--7075 & 23.4 & 1997 Feb 2 &  600 \\
G780H & 6270--8500 & 5.26 & 1997 Feb 2 & 1720 \\
\enddata
\end{deluxetable}

\begin{deluxetable}{ccccccr}
\tablecaption{FOC f/96 Observation Log.\label{tab:obs_log}}
\tablehead{ \colhead{COSTAR} & \colhead{Observing Mode} &
\colhead{Filter} & \colhead{FoV} & \colhead{Scale} & \colhead{Date} & \colhead{Exp.} \\
& & & & \colhead{(\arcsec\ pixel$^{-1}$)} & & \colhead{(s)} }
\startdata
No & 512z$\times$1024 & F175W & 22\arcsec$\times$22\arcsec & 0.0225 & 1993 Nov 5 & 2793\tablenotemark{a} \\
No & 512z$\times$1024 & F220W & 22\arcsec$\times$22\arcsec & 0.0225 & 1993 Nov 5 & 1077 \\
No & 512z$\times$1024 & F275W & 22\arcsec$\times$22\arcsec & 0.0225 & 1993 Nov 5 & 597 \\
No & 512z$\times$1024 & F342W & 22\arcsec$\times$22\arcsec & 0.0225 & 1993 Nov 5 & 417 \\
Yes & 512$\times$512 &  F175W & 7\arcsec$\times$7\arcsec & 0.01435 & 1997 Feb 1 & 1134 \\
Yes & 512$\times$512 & F275W+F1ND & 7\arcsec$\times$7\arcsec & 0.01435 & 1997 Feb 1 & 546 \\
Yes & 512$\times$512 & F342W+F2ND & 7\arcsec$\times$7\arcsec & 0.01435 & 1997 Feb 1 & 396 \\
\enddata
\tablenotetext{a}{Sum of two equal exposures}
\end{deluxetable}

\begin{deluxetable}{clccc}
\scriptsize \tablecaption{Line Emission Fluxes and Modeling
Parameters\label{tab:line_modeling}} \tablehead{ \colhead{Spectrum} &
\colhead{Line} & \colhead{Line Flux} & \colhead{FWHM}
& \colhead{LOS v} \\
\colhead{(1)} & \colhead{(2)} & \colhead{(3)} & \colhead{(4)} &
\colhead{(5)} }
\startdata
G270H & \cii\ $\lambda$2326 & 21.9$\pm$2.4\tablenotemark{a} &
770$\pm$70 & 390$\pm$30 \\
G780H & \oi\ $\lambda$6300 & 43.9$\pm$6.3 & 480$\pm$11\tablenotemark{b}
&
575.6$\pm$5.0\tablenotemark{b} \\
'' & \nii\ $\lambda$6548 & 144.7$\pm$2.4 & '' & '' \\
'' & \nii\ $\lambda$6584 & 434.0$\pm$7.2\tablenotemark{c} & '' & '' \\
'' & \ha\ & 118.0$\pm$6.4 & '' & '' \\
'' & \sii\ $\lambda$6716 & 56.1$\pm$6.5 & '' & '' \\
'' & \sii\ $\lambda$6731 & 62.2$\pm$6.5 & '' & '' \\
\enddata
\tablenotetext{a}{This line is actually a multiplet. Since the
components are unresolved we have modeled it with a single Gaussian
component.}
\tablenotetext{b}{The FWHM and redshift have been
constrained to be the same for all lines in the G780H grating.}
\tablenotetext{c}{The ratio of the two \nii\ lines has been forced to be
exactly equal to the theoretical value 3.0.} \tablecomments{Col.~(1):
FOS grating used in the observations; Col.~(2): Identification of the
emission line; Col.~(3): Integrated flux in the emission line in units
of 10$^{-16}$ erg s$^{-1}$ cm $^{-2}$; Col.~(4): Measured Gaussian FWHM
($\sigma\sqrt{8\ln2}$) in units of km s$^{-1}$; Col.~(5): Observed
redshift in km s$^{-1}$.}
\end{deluxetable}

\end{document}